 \documentclass[final,5p,times,twocolumn,numbers]{elsarticle}

\usepackage{amssymb}
\usepackage{lipsum}
\journal{Annals of Physics}

\begin{document}

\begin{frontmatter}

%% Title, authors and addresses

%% use the tnoteref command within \title for footnotes;
%% use the tnotetext command for theassociated footnote;
%% use the fnref command within \author or \affiliation for footnotes;
%% use the fntext command for theassociated footnote;
%% use the corref command within \author for corresponding author footnotes;
%% use the cortext command for theassociated footnote;
%% use the ead command for the email address,
%% and the form \ead[url] for the home page:
%% \title{Title\tnoteref{label1}}
%% \tnotetext[label1]{}
%% \author{Name\corref{cor1}\fnref{label2}}
%% \ead{email address}
%% \ead[url]{home page}
%% \fntext[label2]{}
%% \cortext[cor1]{}
%% \affiliation{organization={},
%%            addressline={}, 
%%            city={},
%%            postcode={}, 
%%            state={},
%%            country={}}
%% \fntext[label3]{}

\title{Cumulative X-ray Damage in Bismuth Selenide Examined by Simultaneous TXM and XRD}

%% use optional labels to link authors explicitly to addresses:
%% \author[label1,label2]{}
%% \affiliation[label1]{organization={},
%%             addressline={},
%%             city={},
%%             postcode={},
%%             state={},
%%             country={}}
%%
%% \affiliation[label2]{organization={},
%%             addressline={},
%%             city={},
%%             postcode={},
%%             state={},
%%             country={}}

\author[first,second,third]{Sophie E Parsons}
\author[fourth]{Bernard Kozioziemski}
\author[seventh,eleventeenth]{Daewoong Nam}
\author[fourth]{Eric Folsom}
\author[tenth]{Can Yildirim}
\author[eleventh]{Sean Breckling}
\author[twelfth]{Sungwook Choi}
\author[second]{Eric C. Galtier}
\author[eleventh]{Arnulfo Gonzalez}
\author[first,second,third]{Deja Dominguez}
\author[first,second,third]{Emlyn Frederick}
\author[eleventh]{Marylesa M. Howard}
\author[first,second,third]{Sara Jessica Irvine}
\author[first,second,third]{Kento Katagiri}
\author[seventh]{Sangsoo Kim}
\author[seventh]{Seonghan Kim}
\author[seventh]{Sunam Kim}
\author[fourteenth]{Stephan Kuschel}
\author[fifteenth]{R. Stewart McWilliams}
\author[seventeenth]{Norimasa Ozaki}
\author[fourth]{Alison M. Saunders}
\author[twelfth]{Hyunjung Kim}
\author[fourth]{Jon Eggert}
\author[first,second,third]{Leora Dresselhaus-Marais}

\affiliation[first]{organization={Stanford University},%Department and Organization
            addressline={476 Lomita Mall}, 
            city={Stanford},
            postcode={94304}, 
            state={CA},
            country={USA}}
\affiliation[second]{organization={SLAC National Accelerator Laboratory},%Department and Organization
            addressline={2575 Sand Hill Road}, 
            city={Menlo Park},
            postcode={94025}, 
            state={CA},
            country={USA}}
\affiliation[third]{organization={PULSE Institute},%Department and Organization
            addressline={2575 Sand Hill Road}, 
            city={Menlo Park},
            postcode={94025}, 
            state={CA},
            country={USA}}
\affiliation[fourth]{organization={Lawrence Livermore National Laboratory},%Department and Organization
            addressline={ 7000 East Ave.}, 
            city={Livermore},
            postcode={94550}, 
            state={CA},
            country={USA}}
\affiliation[fifth]{organization={Technical University of Denmark},%Department and Organization
            addressline={Anker Engelunds Vej 1}, 
            city={Kongens Lyngby},
            postcode={2800}, 
            country={Denmark}}

\affiliation[seventh]{organization={XFEL Beamline Department, Pohang Accelerator Laboratory},%Department and Organization
            addressline={80, 127-gil, Jigok-ro, Nam-gu}, 
            city={Pohang},
            postcode={37673}, 
            state={Kyungbuk},
            country={Republic of Korea}}
\affiliation[eleventeenth]{organization={Photon Science Center, Pohang University of Science and Technology},%Department and Organization
            addressline={77, Cheongam-ro, Nam-gu  }, 
            city={Pohang},
            postcode={37673}, 
            state={Gyeongbuk},
            country={Republic of Korea}}
\affiliation[eighth]{organization={Danish Technological Institute
},%Department and Organization
            addressline={Gregersensvej 1}, 
            city={Taastrup},
            postcode={2630}, 
            country={Denmark}}
\affiliation[ninth]{organization={Massachusetts Institute of Technology 
},%Department and Organization
            addressline={77 Massachusetts Avenue}, 
            city={Cambridge},
            postcode={02139}, 
            state={MA},
            country={USA}}
\affiliation[tenth]{organization={European Radiation Synchrotron Facility 
},%Department and Organization
            addressline={71 avenue des Martyrs}, 
            city={Grenoble},
            postcode={38000},
            country={France}}
\affiliation[eleventh]{organization={Nevada National Security Site 
},%Department and Organization
            addressline={232 Energy Way}, 
            city={North Las Vegas},
            postcode={89030}, 
            state={NV},
            country={USA}}
\affiliation[twelfth]{organization={Center for Ultrafast Phase Tranformation, Department of Physics, Sogang University 
},%Department and Organization
            addressline={35 Baekbeom-ro, Mapo-gu}, 
            city={Seoul},
            postcode={04107},
            country={Republic of Korea}}
\affiliation[fourteenth]{organization={Technische Universität Darmstadt 
},%Department and Organization
            addressline={Karolinenplatz 5}, 
            city={Darmstadt},
            postcode={64289}, 
            country={Germany}}
\affiliation[fifteenth]{organization={University of Edinburgh
},%Department and Organization
            addressline={Old College, South Bridge}, 
            city={Edinburgh},
            postcode={EH8 9YL}, 
            country={United Kingdom}}

\affiliation[seventeenth]{organization={University of Osaka
},%Department and Organization
            addressline={2-1 Yamadaoka, Suita}, 
            city={Osaka},
            postcode={565-0871}, 
            country={Japan}}
\affiliation[eighteenth]{organization={Universität Hamburg
},%Department and Organization
            addressline={Luruper Chaussee 149}, 
            city={Hamburg},
            postcode={22761}, 
            country={Germany}}

\begin{abstract}
Bismuth selenide ($\mathrm{Bi_2Se_3}$) is a topological insulator with potential applications in thermoelectrics, spintronics, and optoelectronics. However, its response to radiation remains poorly understood. We investigate cumulative X-ray damage in $\mathrm{Bi_2Se_3}$ using simultaneous transmission X-ray microscopy (TXM) and X-ray diffraction (XRD) at the Pohang Accelerator Laboratory X-Ray Free Electron Laser (PAL-XFEL) over 27,000 successive pulses. We observe distinct damage mechanisms: rapid hole formation via vaporization within 100 pulses, followed by slower grain refinement and material sputtering over thousands of thermal cycles. Williamson-Hall analysis reveals a progressive transformation from single-crystal to nanocrystalline structure, with grain sizes decreasing from micron to nanometer scale. Finite-element modeling confirms that X-rays penetrate 13.47 $\mathrm{\mu}$m, driving local temperatures above 1600 K with subsequent cooling between pulses. Scanning electron microscopy identifies three characteristic morphologies corresponding to different thermal histories: sputter streaks, prismatic crystals, and disordered microcrystals. Our results demonstrate that grain-boundary formation creates a feedback mechanism that accelerates damage in later pulses. This work establishes a methodology for studying radiation damage across multiple length scales and provides insight into topological insulator stability under extreme conditions.
\end{abstract}

%%Graphical abstract
%\begin{graphicalabstract}
%\includegraphics{grabs}
%\end{graphicalabstract}

%%Research highlights
%\begin{highlights}
%\item Research highlight 1
%\item Research highlight 2
%\end{highlights}

\begin{keyword}
X-ray ablation \ Topological material \ Bismuth Selenide ($\mathrm{Bi_2Se_3}$ ) \ Radiation Damage \ X-ray Diffraction (XRD) \ Transmission X-ray Microscopy (TXM)
%% keywords here, in the form: keyword \sep keyword, up to a maximum of 6 keywords

%% PACS codes here, in the form: \PACS code \sep code

%% MSC codes here, in the form: \MSC code \sep code
%% or \MSC[2008] code \sep code (2000 is the default)

\end{keyword}

\end{frontmatter}

%\tableofcontents

%% \linenumbers

%% main text

\section{Introduction}
\label{introduction}

Due to its thermoelectric performance and topological insulator behavior, $\mathrm{Bi_2Se_3}$ has emerged as a candidate material for applications in energy harvesting, electronics/spintronics, optoelectronics, photocatalysis, and biomedical imaging and therapy\cite{Banerjee2025Bi2Se3}\cite{Batool2022Barium}\cite{Das2025Bi2Se3}\cite{Li2021Bi2Se3}\cite{Liu2024Bi2Se3}\cite{Yin2016Bi2Se3}. Bismuth Selenide ($\mathrm{Bi_2Se_3}$) has a local structure that has demonstrated feasibility as an important thermoelectric material. Comprised of heavy elements, $\mathrm{Bi_2Se_3}$ has substantial spin-orbit coupling and a unique electronic structure with metallic surface states that are topologically protected from its insulating bulk(\cite{Zhang2009Topological}\cite{Mishra1997Thermoelectric}). How this electronic state interacts with radiation and how ablation changes the behavior of $\mathrm{Bi_2Se_3}$ remains unexplored. 

The electronic structure of $\mathrm{Bi_2Se_3}$ is that of a 3D topological insulator, exhibiting a combination of insulating bulk states and conducting, spin-polarized surface states. Because of this bonding structure, the local and bulk electronic structure can vary. In $\mathrm{Bi_2Se_3}$, bulk bonding is dominated by covalent and weak‑covalent Bi–Se in a layered rhombohedral structure with weak Van der Waals coupling between quintuple layers\cite{Hasan2020Electronic}. The covalent bonding inside each quintuple layer and weak Van der Waals bonding between layers create a layered bulk semiconductor with a narrow bandgap \cite{Zhang2009Topological}. The bulk bands are insulating, but the broken-bond environment at the outer quintuple layers causes gapless, spin-helical surface states that span this gap\cite{MAZUMDER2021161492}. At the surface, broken coordination and where in the material it is terminated determine whether bonding remains Van der Waal‑like or generates dangling bonds. These surface‑specific bonds create additional surface bands, shift and narrow the effective gap near the surface, and have been shown to host strongly spin‑split, topological dangling‑bond states \cite{Hong2010Ultrathin}. Interfaces that preserve Van der Waal bonding generally leave the topological Dirac surface state largely intact\cite{Chiatti2016}. This surface contrasts with an insulating interior and metallic states on the weakly bonded outer layers defines $\mathrm{Bi_2Se_3}$ as a topological insulator.

This difference in global versus local electronic properties becomes more pronounced when comparing single crystal and polycrystalline $\mathrm{Bi_2Se_3}$. Locally, the inter-grain electronic structure is preserved, even in polycrystalline $\mathrm{Bi_2Se_3}$. However, looking across the bulk of the sample, as it degrades from single crystal (SC) to polycrystalline (PC) globally, grain boundary scattering impacts transport properties across the bulk of the material. Studies have clarified that the carrier concentration, mobility, and coherence of topological transport of $\mathrm{Bi_2Se_3}$ is dependent upon the crystallinity of the material. Single crystal $\mathrm{Bi_2Se_3}$ has been shown to exhibit typical bulk insulating/semiconducting with a Dirac-like metallic surface states arising from spin–orbit–driven $\mathrm{p_{(Bi)}–p_{(Se)}}$ band inversion \cite{Jurczyszyn2020Surface}. Single crystal $\mathrm{Bi_2Se_3}$ has a well‑defined Dirac cone and inverted bands near $\Gamma$, with surface metallic states crossing the bulk gap\cite{Zhang2009Topological}. Studies have shown that this is tied to the local inter-grain properties and quintuple bonding, not long‑range single‑crystal perfection\cite{Hasan2020Electronic}\cite{Mishra1997Thermoelectric}. Upon transition from single crystal to polycrystal, the local electronic structure within each grain of a polycrystal remains similar to that of a single crystal. What differs between single crystal and polycrystalline $\mathrm{Bi_2Se_3}$ is the disorder and defects induced by grain boundaries across the bulk of the sample. Polycrystal $\mathrm{Bi_2Se_3}$ shows a net decrease in transport efficacy compared to single crystal $\mathrm{Bi_2Se_3}$. A study by Sahu et al demonstrated that for polycrystalline $\mathrm{Bi_2Se_3}$ with 20nm crystals there exists a very high carrier concentration ($\mathrm{10^{20} cm^{-2}}$) and very low mobility ($\mathrm{8\ cm^2 V^{-1}s^{-1}}$), characteristic of overall semiconducting behavior \cite{Sahu2018Weak}. This study further demonstrated high disorder and grain boundaries dominate scattering with weak anti-localization attributed mainly to impurity‑dominated 2D conduction rather than clean topological surface channels. They further demonstrated that a decrease in the grain size reduces the phonon mean free path, while increasing the density of metallic states \cite{Sahu2018Weak}.

How topological insulators’ electronic structure impacts their ablation dynamics is not well understood. In an insulator, ablation is typically electrostatic rupture from charge accumulation in dielectrics with limited transport\cite{Medvedev2015Femtosecond}\cite{Rethfeld2010Interaction}\cite{PhysRevB.69.054102}. This is due to the low electron mobility in an insulator compared to a conductor. In a metal, bulk thermalization dominates the ablation due to ballistic electrons that can move easily through the conducting material\cite{PhysRevB.69.054102}\cite{Medvedev2011ShortTime}. This leads to melting and vaporization with later timescale effects including bubbling and cavitation, differing from the ionization without thermalization seen in insulators\cite{Medvedev2020ElectronPhonon}. Since $\mathrm{Bi_2Se_3}$ has a metallic surface layer and bulk insulating layers, it is by definition both. Therefore, predicting the radiation-material interactions with this material and the ablation processes that would dominate is difficult. 

Electromagnetic radiation exhibits wavelength-dependent penetration depths in materials. Whereas ablation dynamics are typically studied in the surface-dominated regime, the response of topological insulators in the bulk penetration regime remains poorly understood. Given that the X-ray penetration depth goes beyond the metallically bonded surface layers into the bulk insulator, the interactions span both the complex electronic structure of the surface state and that of the bulk. One study highlighted the importance of skin depth (the area over which energy is deposited) in material ablation, with the larger penetration depth of X-rays predicted to cause bulk insulator behavior\cite{Lee2011Xray}. Theoretical studies have conversely predicted that in topological insulators, the coulomb explosion type ablation would dominate other mechanisms of ablation, leading to a more rapidly ablating material than in a typical insulator. However, this study failed to predict how the degradation of the topological insulators and the ensuing changes to its electronic structure would impact ablation over many damage cycles \cite{Oksengendler2024Radiation}. In a simplified two-component model consisting of metallic surface states and an insulating bulk, one might anticipate sequential removal of the surface layer to expose the underlying insulator. However, in topological insulators, each ablation cycle regenerates metallic surface states on the newly exposed surface until disorder strength reaches a critical point that drives the material into a non-topological state \cite{Sacksteder2015Modification}. The topologically protected surface states preserve the material's spatial and electronic geometry within grains, while the bulk material demonstrates greater disorder and commensurate decrease in transport. Therefore, the impact and result of this protection remains unclear. This protection is particularly ill-defined at ablation timescales, where surface reformation competes with the electronic redistribution processes. Characterizing the parameter space for systems with forbidden versus allowed band gaps under external fields is therefore essential. Prior work has determined that the bulk bandgap range of undoped $\mathrm{Bi_2Se_3}$ is 0.2-0.3 eV \cite{MAZUMDER2021161492}. However, in thin and nanostructured forms, strong confinement can increase the effective gap up to several eV. Surface band bending and heterostructure engineering introduce additional allowed quantum well states and tunable direct/indirect gaps \cite{Liu2014Tuning}\cite{10.1063/1.4975819}\cite{Dong2026Lattice}. 

While X-rays are typically used to probe material properties, their ability to penetrate deeply into materials also enables volumetric energy deposition as the electromagnetic radiation (EM) interacts with the lattice, causing damage. The mechanisms by which light structurally disrupts a material depend heavily on wavelength and energy. At higher frequency light (X-rays), damage is induced by direct ionization, typically in the form of Auger or photoionization processes. In contrast, optical radiation at the highest intensities (~terawatts) disrupts materials through collisional absorption of EM radiation -- often called inverse bremsstrahlung or Joule heating -- which can rip out electrons and initiate ablation. Following the initial energy transfer from EM radiation to the material, processes such as thermalization, vaporization, structural phase transitions, mechanical removal of material, and atomization occur\cite{Colvin2013Extreme}. Together these are referred to as ablation. The form this ablation takes, the processes that occur, and the timescales over which they proceed depend not only on the characteristics of the EM radiation source, but also on the electronic and structural properties of the absorbing material.

Currently, in the field of ablation, there are limitations on diagnostics capable of observing radiation-material interactions and subsequent ablation dynamics \cite{Sun2024Dynamics}. Ablation in many bulk materials occurs through random, unpredictable events. The stochastic nature of radiation-material interactions makes them difficult to predictably locate and measure with the appropriate time and spatial resolution with pump-probe experiments. As such, there are two prominent methods by which these dynamics are frequently studied. Some studies tightly focus a femtosecond beam to observe local dynamics\cite{Winter2020Ultrafast}\cite{Yao2022Exploring}\cite{Guo2019Ultrafast}. While highly informative, this measurement geometry changes the ablation dynamics due to beam size and pulse duration. Other studies instead take average measurements over bulk dynamics, losing spatial resolution and sensitivity to local variations in dynamics\cite{Parsons2024Ablation}\cite{Burdt2009Scaling}\cite{Joshi2023Observation}. Because the ablation dynamics vary across length scales from micro to bulk effects, full classification of ablation dynamics requires the combination of both these types of techniques simultaneously. This is enabled by X-ray sources.

Prior experiments at synchrotrons have used X-rays to quantify ablation. Building on previous laser damage work conducted at synchrotrons, Park et al conducted a series of experiments using far-field High-Energy Diffraction Microscopy (HEDM) and combined Small and Wide Angle X-Ray Scattering (SAXS/WAXS) to study grain‑resolved strain, texture, and nano‑precipitation in metals \cite{Park2015Synchrotron}. While these experiments were able to quantify the effects of ablation on the material, the ablation dynamics themselves occurred at too rapid a time scale to resolve. To resolve this issue a series of experiments at the synchrotron used single‑pulse phase‑contrast imaging at MHz frame rates captures ablation plasmas driving shocks into foams and solids, resolving elastic compression, pore collapse, fracture and fragmentation in polyurethane and graphite. This MHz frame rate was able to capture late time scale ablative shock effects but could not resolve sub 100 ps time scale and sub 8 $\mathrm{\mu}$m length scale effects \cite{Olbinado2018Ultrahigh}.

Over the last decade, with the advent of X-ray free electron lasers\cite{Kang2017Hard}, the fully coherent ultrafast and ultrabright pulses analogous to a conventional ultrafast laser have ushered in a new frontier of X-ray science, unlocking time and length scales not accessible by synchrotron radiation\cite{Yabashi2017Next}. This offers the opportunity to probe complex systems and ultrafast processes like those found in the ablation directly on nanometer length scales and femtosecond time scales. While powerful in their diagnostic capabilities XFELs exhibit high fluxes that may introduce a wide range of different damage pathways. Typically, the paradigm of “diffract before destruct” is commonly used to image materials before the X-ray damage accumulates \cite{Chapman2014Diffraction}. For example, Parsons et al used XRD to quantify laser-based ablation depth in aluminum on the picosecond (ps) time scale; however, this technique was not sensitive to the mechanism of ablation and averaged over the bulk material\cite{Parsons2024Ablation}. Time‑resolved Grazing-Incidence Small-Angle X-ray Scattering (GISAXS) tracked early stages of femtosecond (fs) ablation of Si, resolving the formation of liquid droplets in the plume with 20ps resolution, giving nanoscale information on nucleation during ablation \cite{Hull2019Early}. Similarly, this technique was also used on multilayer samples to study subsurface density dynamics and surface ablation with nanometer depth resolution\cite{Randolph2022Nanoscale}. This study was able to resolve heat transport, plasma formation, and re-solidification below the surface but was not sensitive to bulk dynamics. Together these previous studies reveal a key gap in the field that remains: it is either possible to quantify bulk dynamics or resolve microscale effects of ablation but a simultaneous measurement across length scales has not yet been achieved.

Our work explores how the ablation of $\mathrm{Bi_2Se_3}$ evolves during successive cycles of nonuniform X-ray heating. Building off of the work presented in Katagiri et al (2025) \cite{Katagiri2025Xray}, we use simultaneous X-ray transmission microscopy (TXM) and classical X-ray diffraction (XRD) at PAL-XFEL to demonstrate localized X-ray beam damage \cite{Kang2017Hard}. While the brightest point in the X-ray beam quickly degrades the sample, the diffuse edges transform the initially single-crystal to successively smaller grain sizes, showing ordered and disordered crystals as well as sputter. The TXM images reveal how the debris and edge structures evolve over successive pulses, while diffraction patterns capture the emergence of the polycrystalline structure. With this information, we can demonstrate how the local and long-range structures indicate different damage mechanisms. With finite-element simulations, we predict how X-ray dose translates into the maximum temperature and phase of the material and sets the local cooling rate. Across 27,000 of thermal cycles, our simultaneous measurements reveal the importance of temperature stability in setting the dominant damage pathways over different timescales of thermal cycling. Our view of $\mathrm{Bi_2Se_3}$ holds opportunities for a wide range of materials, which can now use simultaneous TXM and XRD at XFELs to gain new insight in simultaneous localized and bulk transformations across materials science. 
\section{Experimental Setup}
\begin{figure}
	\centering 
	\includegraphics[width=0.4\textwidth]{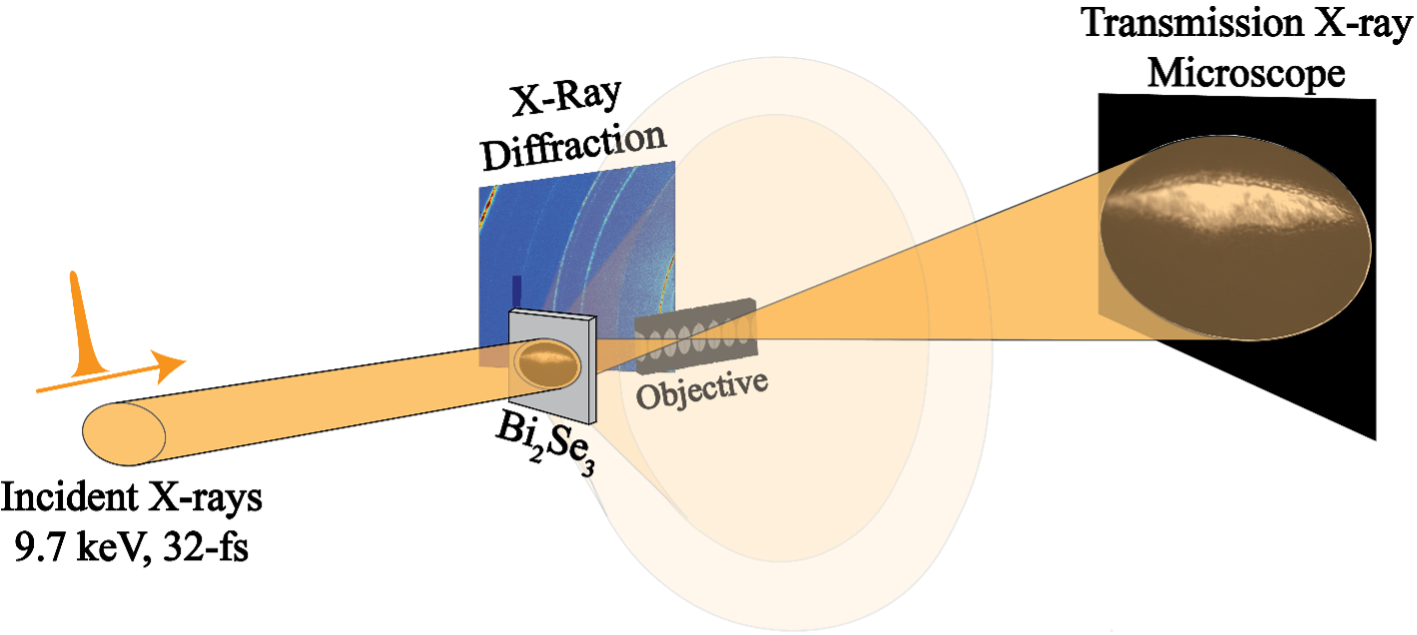}	
	\caption{Schematic showing the setup for TXM measured simultaneously with classical X-ray diffraction (XRD) and optical microscopy} 
	\label{fig_1}%
\end{figure}

This work was conducted at the Nano-crystallography and Coherent Imaging (NCI) of PAL-XFEL \cite{Kang2017Hard}\cite{Kim2018Focusing}. It explores the evolution of grain boundaries based on X-ray heating with successive 32-fs duration pulses at 30-Hz repetition rates at an X-ray energy of 9.7 keV and a fluence of 0.5 mJ per pulse. With a single X-ray pulse hitting the sample, our experiment simultaneously collects the transmitted and diffracted rays to resolve complementary information (Fig. \ref{fig_1}). The transmitted light is collected by the transmission X-ray microscopy (TXM), following the configuration analogous to the synchrotron version. As described in Methods (Figure \ref{fig_1A}), we use an upstream lens to focus the XFEL beam down to the ~15-$\mathrm{\mu}$m diameter oval that defines our field of view, and magnify the image of our sample using compound refractive lenses after the sample, which we collect on the imaging detector, as shown in Fig. 1. We simultaneously collect the wide-angle diffracted X-rays on an area detector to resolve the long-range crystal structure arising from the microscopic structure present in the same region. Using the “diffract-before-destruct" principle [9], our experiment measures the structural changes in the sample caused by the previous pulse before further degrading the material. Our image and XRD sequences thus show snapshots of the progression of the cumulative damage from thermal cycling on each XFEL pulse over ~15 minutes on a 63 $\mathrm{\mu}$m thick sample.

\section{Results}
\begin{figure}
	\centering 
	\includegraphics[width=0.4\textwidth]{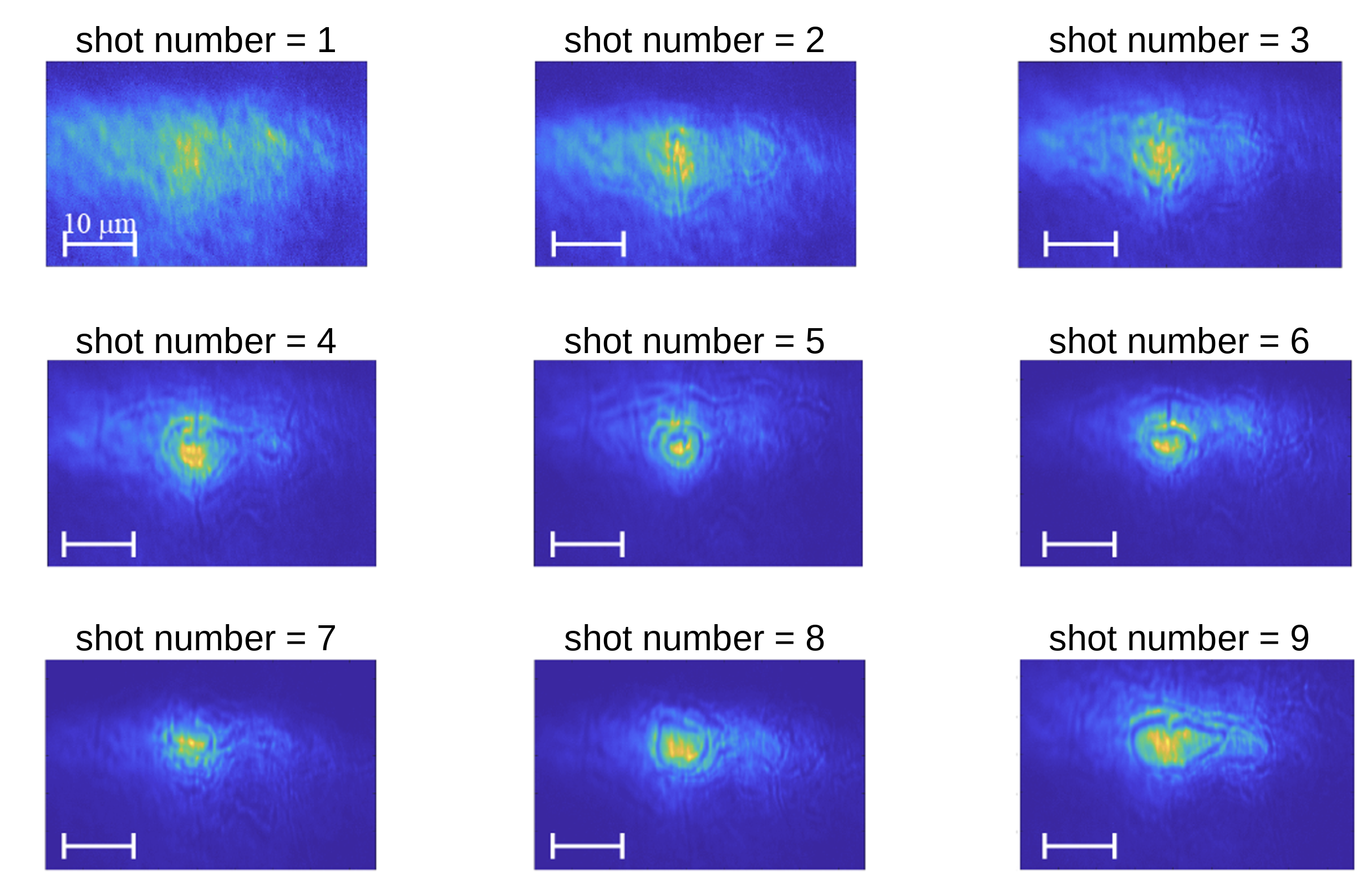}	
	\caption{Initial onset of damage over the first 9 pulses from the XFEL} 
	\label{fig_3}%
\end{figure}

\begin{figure*}[t]
    \centering
    \includegraphics[width=0.85\textwidth]{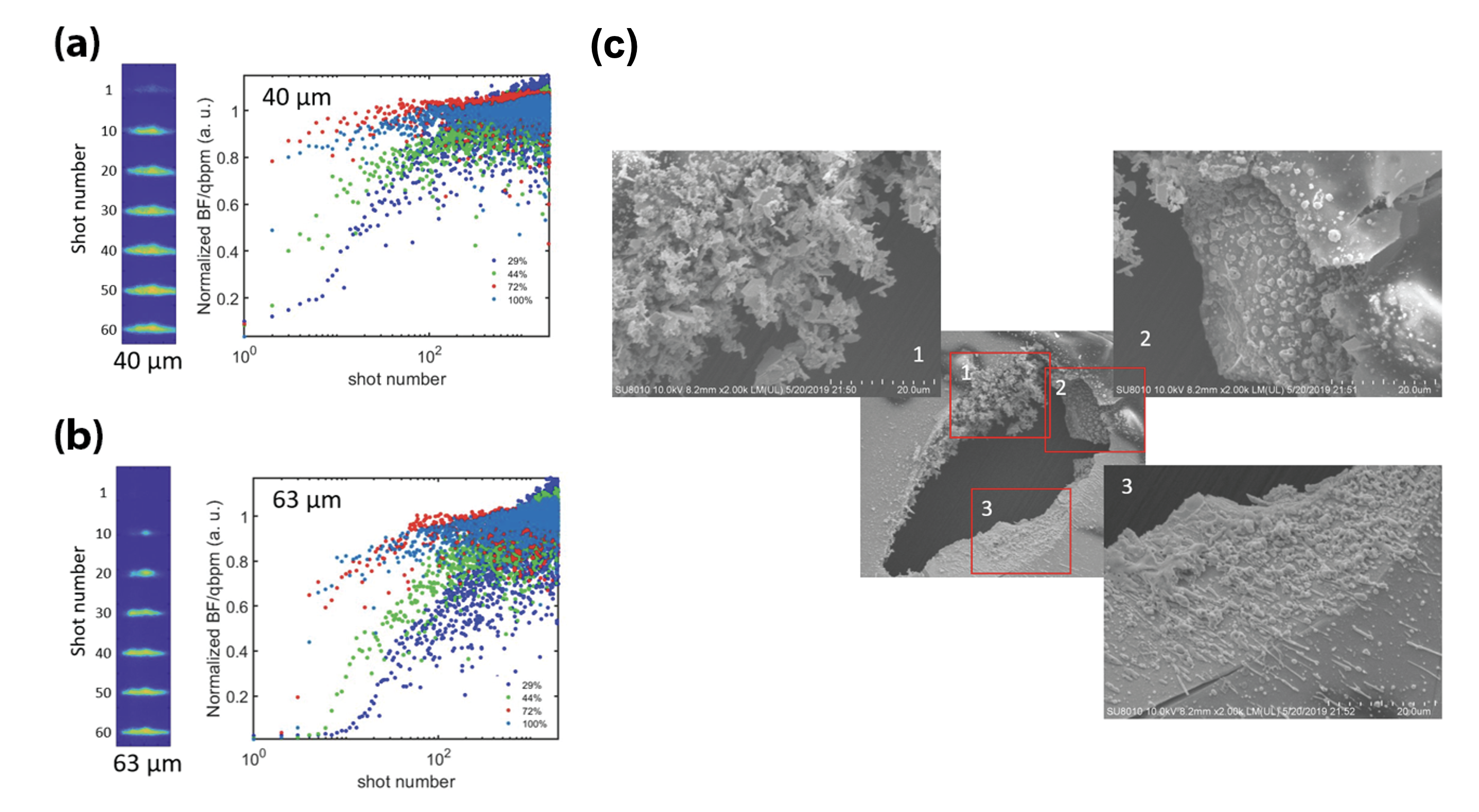}
    \caption{Data showing the onset of radiation damage in Bi$_2$Se$_3$ from the 2019-1st-NCI-037 experiment. Selected images of the 63-$\mathrm{\mu}$m and 40-$\mathrm{\mu}$m thickness samples are shown (29\% fluence) to illustrate the progression of damage as a function of sample thickness. Panels (a) and (b) show the normalized TXM intensity for each XFEL shot, where differences in slope indicate distinct active damage pathways. These results demonstrate that low-flux measurements are required to resolve the complex radiation-damage pathways at this time resolution. Panel (c) shows SEM images illustrating three observed damage mechanisms: sputtering (3), recrystallization into small prismatic crystals (2), and flash recrystallization into disordered crystalline flakes (1).}
    \label{fig:fig3}
\end{figure*}

At 9.7 keV, $\mathrm{Bi_2Se_3}$ has a penetration depth of 13.47 $\mathrm{\mu}$m, as bismuth and selenium both have relatively high absorption cross sections. With strong X-ray attenuation, $\mathrm{Bi_2Se_3}$ has high contrast in TXM when compared to the surrounding air. The initial TXM images (Fig. \ref{fig_3}) demonstrate that the initial pulses formed diffuse structures over the brightest region of the spot, which quickly evolved into a hole that grew larger with each successive XFEL pulse. As the region inside the hole is significantly brighter than the $\mathrm{Bi_2Se_3}$ region, this clearly demonstrates that the XFEL cleared a hole into the material at very early timescales. 
With the strong intensity difference between $\mathrm{Bi_2Se_3}$ and air, we were able to use the integrated intensity collected by TXM (normalized against pulse energy) as a figure of merit to describe how damaged the material was after each successive XFEL pulse. Shown in Fig. \ref{fig:fig3}, these graphs allow us to evaluate the long-term trends in cumulative radiation damage pathways based on changes in the curve slopes. We explored  how the $\mathrm{Bi_2Se_3}$ damage mechanism varies with sample thickness and X-ray fluence. We observed that the XFEL forms a hole in the initial crystal within 100 shots at every XFEL flux, but the behavior continues slowly over thousands of subsequent shots. 

As high-intensity X-ray heating has been studied since the advent of XFELs \cite{AlonsoMori2020Femtosecond}\cite{Tachibana2015Nanoplasma}\cite{Sawada2024Spatiotemporal}, we used thermal models to understand the flow of energy from the X-ray pulse to the heat and forces in the sample that impart the initial damage. X-rays are attenuated according to Beer’s law, with the most energy deposited at the incident surface and decreasing with depth into the sample.  The 1/e depth for $\mathrm{Bi_2Se_3}$ at 9.7 keV is 13.5 $\mathrm{\mu}$m, smaller than the thickness of the samples (40 $\mathrm{\mu}$m and 63 $\mathrm{\mu}$m) used in this experiment, which introduces a thermal gradient within the sample.  

We use a simplified COMSOL Multiphysics finite-element model to simulate heat transfer and phase transitions induced by the XFEL pulse (refer to Methods: COMSOL model for more details). and the flow of heat within the sample between pulses. Figure \ref{fig_2A} shows the resulting temporal profile for multiple positions within the sample. Over the first 0.5 ns, the energy is applied as a uniform pulse in time. The XFEL beam is incident on the sample at 0 $\mathrm{\mu}$m and is subjected to the most energy.  In the simulation, at 980 K the phase changes from solid-to-liquid and  at 1600K the phase changes liquid-to-vapor. The thermal diffusivity is low in the solid and liquid, so there is little transport of heat before 0.1 µs. The heat is transported both through the thickness and radially outward on time scales of 1-100 µs.  The sample would be expected to recondense and solidify by approximately 1 ms, and back down to nearly room temperature by the next 33 ms pulse. 

As mass transport is not included, the portion of the sample that is expected to vaporize is not accurately captured and there is likely more energy in the model at late times than would be expected in experiment.  The model demonstrates that approximately the first 16-20 $\mathrm{\mu}$m of the sample would be expected to vaporize, and melting is expected down to approximately 30-35 $\mathrm{\mu}$m into the sample.  We can expect that some fraction of the vaporized material will be removed from the sample as the XFEL drills through the sample. The second XFEL pulse, then, will see approximately 10-20 $\mathrm{\mu}$m of material that has been melted and re-solidified, and some material that will have vaporized and re-condensed. We would expect that the 40 $\mathrm{\mu}$m and 63 $\mathrm{\mu}$m thick samples are drilled through in 3-4 pulses and 4-5 pulses respectively, assuming that the beam position and intensity does not substantially change from shot to shot. This is consistent with measurement of the transmission through the sample as shown in Figure \ref{fig_3}.  

With the thermal model in hand, we can explore the pulse-to-pulse energy variation using the Quadrant Beam Position Monitor (QBPM) data and the pulse-to-pulse position changes on the sample morphology.  We acquired a series of images without a sample to quantify the statistical changes in the beam. We use the imaged beam to create an intensity map of the beam on the sample, with the integrated intensity equal to the total pulse energy. Next, we scale the amount of material removed based on the local intensity variations M(x,y,n), where n is the pulse number.  The remaining thickness is then available for the next shot, with the new intensity F(x,y,n+1).  Finally, we can compute the transmitted intensity for each pulse based on the material remaining. Comparison between the experimentally measured transmitted intensity and the simulation shows good agreement for the first 100 pulses in both the 40 $\mathrm{\mu}$m and 63 $\mathrm{\mu}$m thick samples when we use 12 $\mathrm{\mu}$m of material removed for a flux of 0.25 µJ/$\mathrm{\mu}$m2. This is a little lower than the thermal model which suggests 16-20 $\mathrm{\mu}$m is vaporized for the same flux.  

Beyond the initial heating process, these models also demonstrate that the heat imparted by the initial XFEL pulse heats first along the original optical path over a 10 $\mathrm{\mu}$s timescale, then diffuses radially over a ms timescale, until it returns to sub 500K by the time that the subsequent XFEL pulse reaches the sample 33-ms later. The models also demonstrate that thermally-driven changes to the material span 50 $\mathrm{\mu}$m out from the initially irradiated region, explaining some of the features that appear in Figure \ref{fig_3}. 

Beyond the first 100 pulses, our XRD results indicate that transformations continue for all 27,000 pulses from the XFEL. To demonstrate this, we show the imaging and XRD results side by side in Figure \ref{fig:fig3}a,b, demonstrating the continued damage trends we resolve over longer timescales. We plot the integrated TXM intensity as a function of XFEL shot to show the different timescales over which discrete damage mechanisms occur in the material. Changes to the slopes of the curves indicate that the lowest intensity XFEL fluences have more detailed damage mechanisms.

\begin{figure*}[t]
    \centering
    \includegraphics[width=0.85\textwidth]{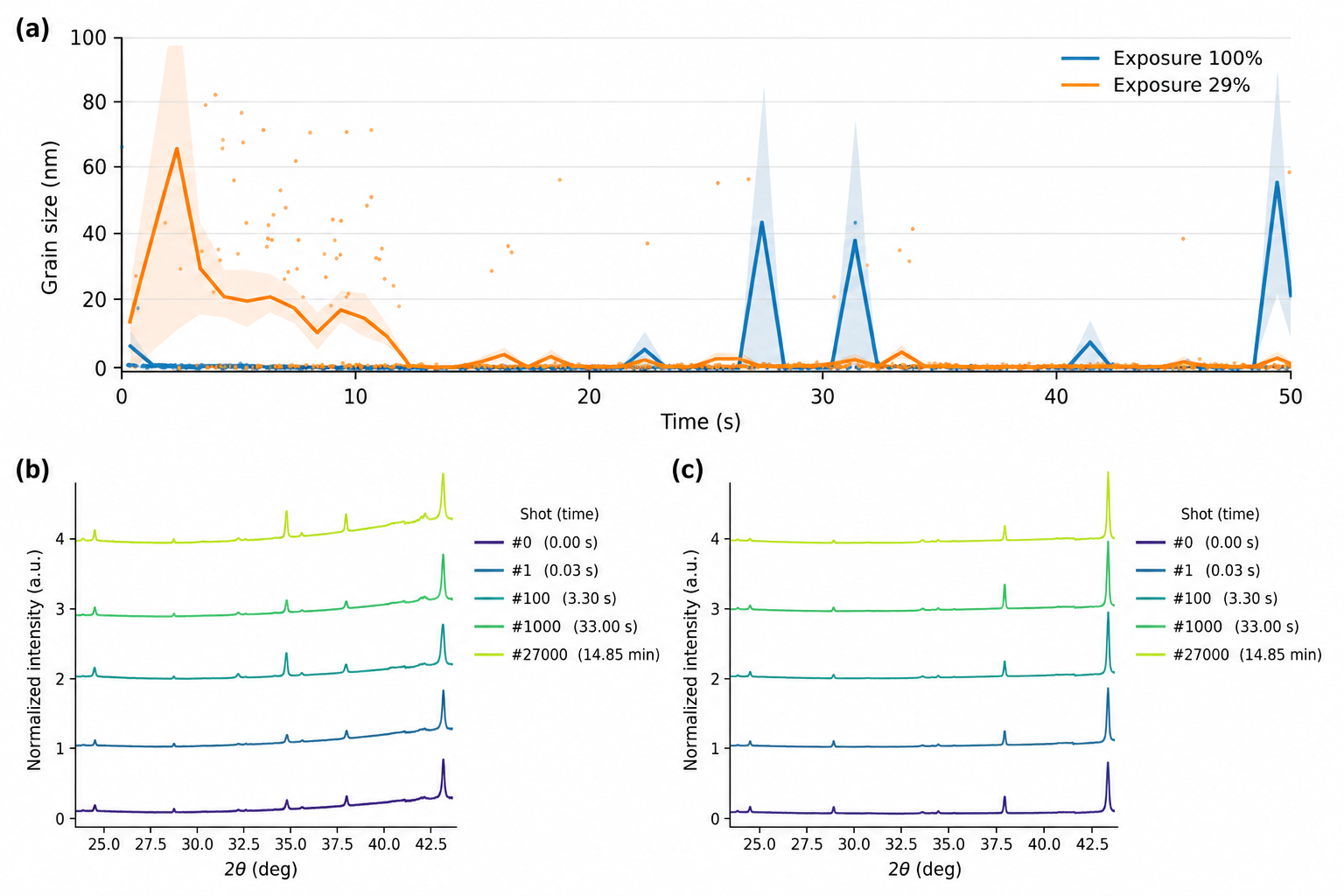}
    \caption{a. Williamson-hall analysis of grain size as a function of pulse duration with X-ray exposures of 100\% and 29\% b. Raw XRD trace for pulse numbers 0,1,100,1000 and 27000 for an X-ray exposure of 100\%b. c. Raw XRD trace for pulse numbers 0,1,100,1000 and 27000 for an X-ray exposure of 29\% }
    \label{fig:fig4}
\end{figure*}

During the X-ray ablation processes in the $\mathrm{Bi_2Se_3}$ we do not observe amorphization under the solid peaks, nor do we observe a left shift in the diffraction peaks, which would be characteristic of heating in the system. Using the COMSOL model (Fig. \ref{fig_2A}), this is to be expected as the "diffract before destruct" principle used here means we are visualizing the damage caused by the previous cycle, and melting and vaporization occur on longer time scales than are captured by each XFEL pulse. Together, this confirms the re-solidification between X-ray pulses as predicted in our thermal model (Fig. \ref{fig:fig4}). 

Scanning electron microscopy (SEM) of the recovered samples (Fig. \ref{fig:fig3}c) revealed three different types of damage: streaks of amorphous material typical of sputter (3), small prismatic crystals (2), and small clusters of micro-grains (1). Image analysis using ImageJ finds that the average grain size in (1) range from approximately 5$\mathrm{\mu}$m to 1$\mathrm{\mu}$m, in (2) range from approximately 3$\mathrm{\mu}$m to 0.1$\mathrm{\mu}$m, in (3) the sputter length increases as a function of distance from crater. At the lip of the crater, the sputter marks are near continuous for 13$\mathrm{\mu}$m. It then increases in length and decreases in volume as a function of distance from the crater, with a final average length of 5$\mathrm{\mu}$m. 

We find in Figure \ref{fig_3} that many of the initial images show vertical lines emanating from the hole in the sample as it grows. Using SEM, we are able to connect these lines to potential damage pathways in our material. In the SEM image, the streaks of amorphous material below the damaged spot are typical of the patterns one would see from a material that sputters, suggesting that the material transformations initiated by the XFEL include melting. The sputter pattern suggests that the initially ~20 °C material reached a sufficiently high temperature to reach the melt, which occurs at ~710 °C under equilibrium conditions. We therefore interpret the vertical lines in the TXM image as melted material being ejected from the sample. 

We additionally see the formation of both organized and disorganized crystal growth upon recrystallization. Assuming an initial adiabatic melt initiated the radiation damage, the other damaged regions would then arise from different thermal gradients upon cooling which are likely from either different $T_{\mathrm{max}}$ values, or diffusion conditions. 

\begin{figure*}[t]
    \centering
    \includegraphics[width=0.85\textwidth]{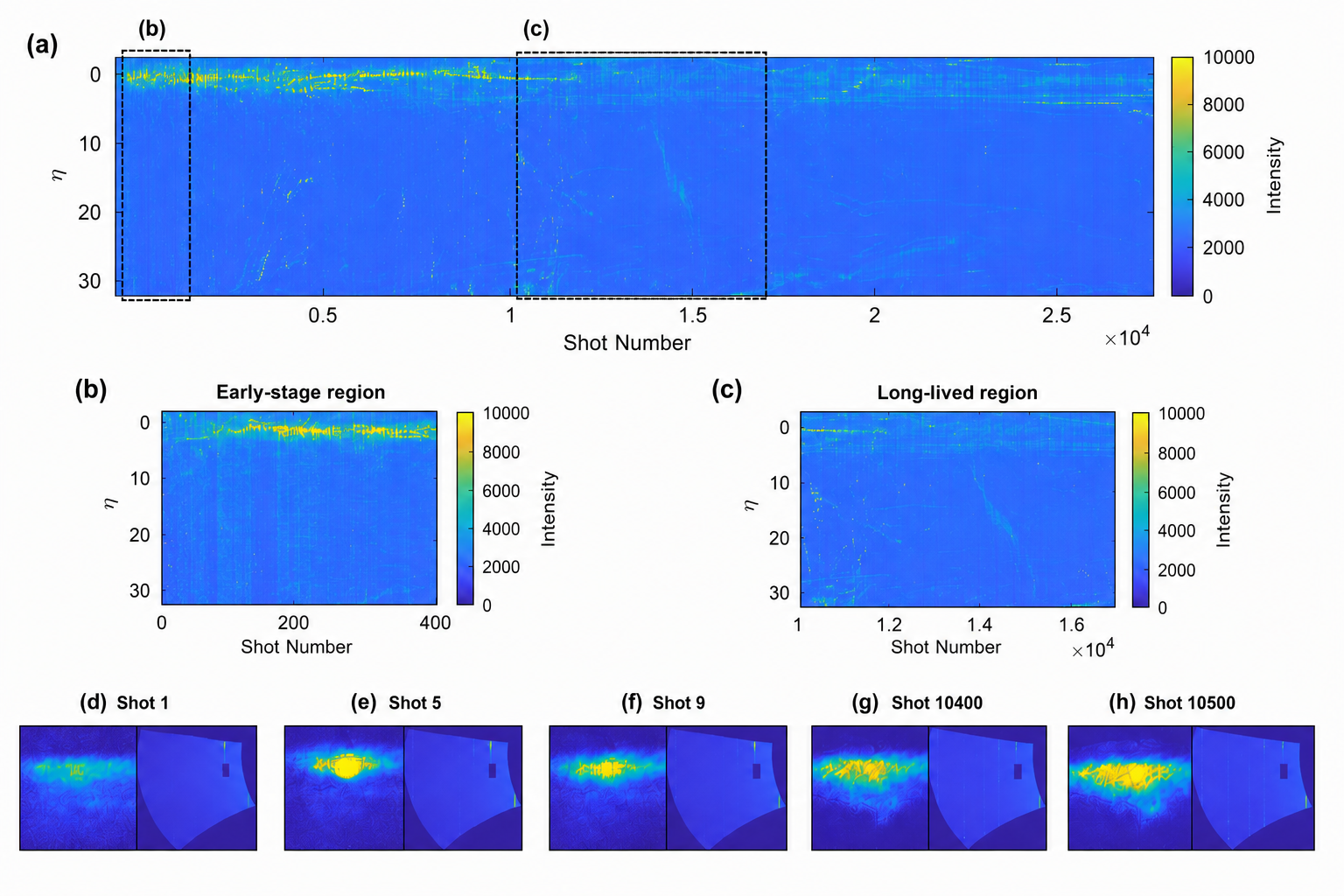}
    \caption{a. Raw XRD data of the peak located at 2$\theta$=35.7$^{\circ}$ as a function of shot number. The highlighted regions correspond to zoomed in regions b and c. b-c. Raw XRD data of the peak located at 2$\theta$=35.7$^{\circ}$ for b: the first 400 X-ray pulses and c: X-ray pulses number 10000 to 17000. d-g. TXM and full raw XRD patterns for  pulse numbered d: 1. e: 5 f: 9 g: 10400 h: 10500 }
    \label{fig:fig5}
\end{figure*}

\begin{figure*}[t]
	\centering 
	\includegraphics[width=0.8\textwidth]{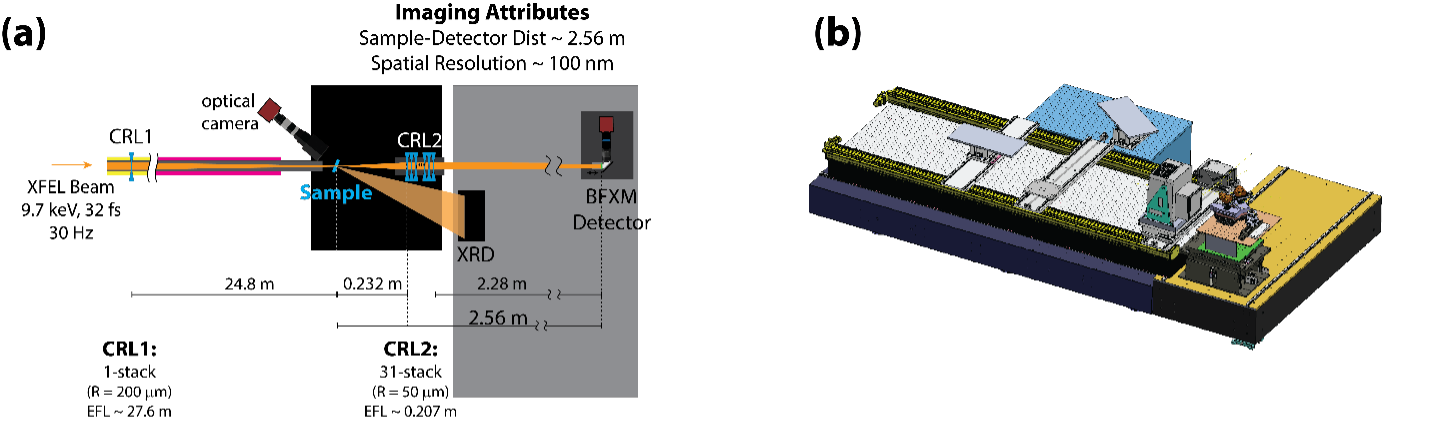}	
	\caption{(a) Schematic showing the setup for TXM measured simultaneously with classical X-ray diffraction (XRD) and optical microscopy, with (b) a 3D model of how the experimental configuration fit into the NCI hutch.} 
	\label{fig_1A}%
\end{figure*}

\begin{figure}
	\centering 
	\includegraphics[width=0.4\textwidth]{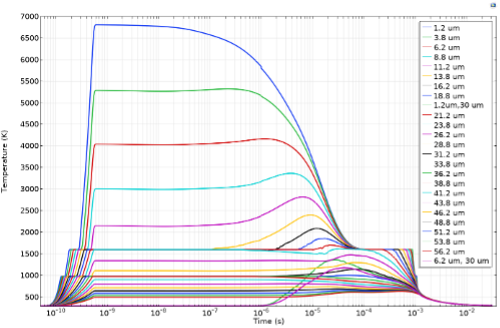}	
	\caption{Sample temperature as a function of time after the XFEL pulse.  Each line is a different location within the sample, with the XFEL beam incident at 0 $\mathrm{\mu}$m.  The steps visible at 980 K and 1600 K are the locations of the solid-to-liquid and liquid-to-vapor transitions. } 
	\label{fig_2A}%
\end{figure}

We observe the transition of the initially single crystalline material to polycrystalline in our diffraction as well. Our in‑situ diffraction shows a transition from single‑crystal diffraction to a spotty ring indicative of powder-adjacent diffraction scattering under repeated XFEL irradiation (Fig. \ref{fig:fig5}d-h). Our Williamson–Hall (WH) analysis of the in-situ XRD quantifies this evolution as a progressive reduction in crystallite size with a concurrent change in microstrain. At 100\% transmission we see a rapid decrease in crystallite size as a function of time with grains after cycle 1 on the order of sub 10nm. Conversely at 20\% exposure, the grains decrease from an order of 100nm to sub 10nm over repeated irradiation cycles. This implies that the speed over which crystallization occurs may depend on the intensity of radiation. Some outliers exist in the data due to inherent fitting errors as captured by the shaded region around the lines in figure \ref{fig:fig4}a. 

While the X-ray diffraction patterns remain consistent in broadness and 2$\theta$ throughout the ablation process (Fig. \ref{fig:fig4}), zooming in on the ring located at 2$\theta$=35.7$^{\circ}$, we find a change in the diffraction pattern in $\eta$ as a function of pulse number (Fig. \ref{fig:fig5}). 

The fast initial damage, as confirmed through TXM (Fig. \ref{fig:fig5}b, d-f) bores a hole through sample, while the longer time scale damage emerges at later times (Fig. \ref{fig:fig5}c,g,h) as the diffuse edges of beam further erode the edges of hole.

\section{Discussion}

Interpreting the WH crystallite size as a proxy for the characteristic spacing between grain boundaries, we see evidence in both our raw diffraction pattern and the WH analysis that the X-ray irradiation drives $\mathrm{Bi_2Se_3}$ from a single crystal to a low‑boundary‑density state to a nanocrystalline, boundary‑dominated microstructure. We see that this transition varies as a function of X-ray intensity, with higher X-ray flux driving a faster transition to nanopolycrystallinity. 

The SEM and WH analyses resolve ablation at complementary length scales with SEM revealing micron-scale morphology and WH resolving grain sizes down to the nanometer scale. Together with TXM, these diam=gnostics confirm that the transformation proceeds via two concurrent damage pathways operating on different timescales (Fig. \ref{fig_3}). There is rapid, localized material removal at the beam center linked to the intense core of the X-ray pulse, and slower, progressive grain refinement in the surrounding material linked to cumulative thermal cycling from the diffuse beam edges. Bulk diffraction patterns remain consistent in broadness and $2\theta$ (Fig. \ref{fig:fig4}b,c) even as individual rings sharpen and redistribute in $\eta$ (Fig. \ref{fig:fig5}a-c) which confirms that this second, slower pathway is a boundary-density effect, not a bulk phase or lattice-parameter change.

This grain refinement links our diffraction results to expected changes in electronic transport. As WH-derived crystallite size decreases, grain-boundary density increases. Prior studies of polycrystalline and nanopolycrystalline $\mathrm{Bi_2Se_3}$ consistently report that higher boundary density brings increased carrier concentration and reduced mobility relative to single crystals, driven by boundary- and defect-assisted doping and scattering \cite{Sacksteder2015Modification}\cite{Chiatti2016} \cite{Hasan2020Electronic}\cite{Mishra1997Thermoelectric}. Therefore, the WH trend implies that repeated irradiation cycles should progressively suppress coherent, high-mobility topological transport and shift the material toward disordered, boundary-limited conduction, even if the local band inversion within individual grains remains largely intact. These studies of polycrystaline $\mathrm{Bi_2Se_3}$ have demonstrated that the same grain boundaries that degrade transport also increase electron-phonon coupling and accelerate lattice heating. We would expect therefore the emergence of polycrystaline  $\mathrm{Bi_2Se_3}$ upon repeated ablation cycles to have a lower local thermal threshold for further melting and ablation compared to the pristine single crystal. The preferential orientation of the resultant grains is described in Katagiri et al (2024)\cite{Katagiri2025Xray}. 

Our XRD, TXM and SEM together demonstrate that irradiation refines the grains. These refined grains are expected to couple X-ray-deposited energy into heat more efficiently, and this enhanced heating would then drive further grain refinement and material loss in subsequent pulses. The grain-boundary evolution seen in SEM, WH, and XRD is the mechanism that accelerates the damage rate itself, and it is likely the reason the slope of the damage curves in Fig. \ref{fig_3} change over pulse cycle.

\section{Conclusion}

X-ray irradiation transforms $\mathrm{Bi_2Se_3}$ through sequential mechanisms. Initially ablation occurs as a rapid, localized vaporization and hole-drilling at the beam center within the first 100 pulses, driven by direct thermal melting/vaporization. After the initial hole is drilled, ablation continues as a cumulative transformation of the surrounding single-crystal material into a nanocrystalline, grain-boundary-dominated microstructure over the full 27,000 pulses. Our Williamson-Hall and SEM analysis shows that repeated thermal cycling progressively refines the crystallite size from micron to nanometer scale, increasing grain-boundary density with each cycle. Because grain boundaries both scatter carriers and enhance electron-phonon coupling, this structural evolution is expected to simultaneously degrade the coherent, high-mobility topological surface transport that motivates  $\mathrm{Bi_2Se_3}$’s applications, and lower the local thermal damage threshold, meaning each successive pulse causes disproportionately more structural damage than the last. Ablation does not just remove material from $\mathrm{Bi_2Se_3}$, it also progressively converts the topologically-protected single crystal into a disordered, boundary-limited polycrystal with transport and thermal properties expected to be governed by defect scattering rather than the protected surface state, even if the local band inversion within individual grains persists.

Beyond $\mathrm{Bi_2Se_3}$, our demonstration of simultaneous TXM and XRD at XFEL repetition rates establishes a general methodology for studying dose-dependent transformations where local morphology and long-range order evolve on disparate timescales. The ability to track hole drilling, phase transitions, and grain refinement in the same field of view resolves a longstanding diagnostic gap in radiation damage studies. This approach is applicable with applications ranging from radiation-hard materials for fusion and fission reactors, pulsed-laser deposition dynamics, shock-induced phase transitions in high-energy-density physics, and localized thermochemical reactions in catalysis and battery electrode degradation. For topological materials specifically, our results suggest that the surface-state coherence time under intense X-ray pulses is limited not by direct ionization (which would favor Coulomb explosion) but by the accumulated grain-boundary disorder from repeated thermal cycling. The agreement between COMSOL thermal predictions, TXM-derived hole-drilling rates, and WH grain-size evolution validates thermal modeling as a predictive tool for XFEL sample design, enabling optimization of dose, spot size, and repetition rate to maximize signal while controlling damage for future in-situ studies of functional materials under extreme conditions.

%%\label{}

\section*{Acknowledgments}
Stanford/SLAC work from this study was supported by the Department of Energy, Office of Science, Basic Energy Sciences, Materials Sciences and Engineering Division, under Contract No. DE-AC02-76SF00515. B. K., E. F., A.S, and J.E. acknowledge their work under the auspices of the US Department of Energy by Lawrence Livermore National Laboratory under Contract No. DE-AC52-07NA27344. These experiments were performed using the NCI instrument at PAL-XFEL (Proposal No. 2020-2nd-NCI-028) funded by the Ministry of Science and ICT of Korea. S.C., S. K. and H.K. acknowledge the support from the National Research Foundation of Korea (RS-2021-NR059920).

\section*{Methods}
\subsection*{Simultaneous TXM/XRD}
These experiments were performed at the PAL-XFEL, using the NCI experimental hall (EH2). We used the QBPM and in-line spectrometer in the Optics Hall to calibrate the pulse energy and spectrum, respectively, for each X-ray pulse. While the variation was small, this type of internal calibration helped guide some of our conclusions, as the self-amplification of stimulated emission (SASE) process can cause significant spectral and amplification jitter between successive pulses. The QBPM instrument measured the voltage generated from capacitors in the X-ray beam, which were converted to the X-ray pulse energy using the 9.7 keV calibration that was collected previously. 

Downstream of the X-ray pulse calibration measurements, we installed a custom optical setup into the NCI endstation to simultaneously measure TXM and XRD images on each pulse of the XFEL. Classical XRD was measured on a Rayonix MX225-HS detector, which was placed 444 mm from the sample, with the detector normal to the axis of propagation of the incident XFEL. With the 4096 x 4096 array of ~15-$\mathrm{\mu}m$ pixels, our data collected XRD covering a range of $22^\circ \leq 2\theta \leq 45^\circ$ with the azimuthal range of $-13.5^\circ \leq \theta \leq 45.6^\circ$. To eliminate issues with data transfer and file sizes, we collected XRD images using 2x binning.

For TXM, we placed a stack of CRLs between the sample and detector to image the material—similar to a classical optical microscope. The ~50-$\mathrm{\mu}m$ field of view in our images corresponds to the X-ray spot size at the sample, set by an upstream 1-lens CRL (R = 300-$\mathrm{\mu}m$) that was placed 23.3 m before sample. We then placed a stack of 31 Be CRLs (R = 50-$\mathrm{\mu}m$) 24 cm behind the sample, with a scintillator screen placed another 2.6 m along the transmitted beam, producing a 9.4x X-ray magnification. The 1-cm diameter 35-$\mathrm{\mu}m$ thick Ce:LuAG screen then converted the ~1-mm diameter X-ray beam to its optical analogue, passing through a 7.5x Mitutoyo infinity corrected objective and f = 200-mm lens to relay-image the TXM image onto the Manta 046B detector (8.3-$\mathrm{\mu}m$ pixels). The additional optical magnification afforded our images an overall 70.5x magnification, which nears the maximum resolution attainable with the aberrations of the parabolic CRLs. 
\subsection*{COMSOL model}
We start by assuming a flat-top beam with uniform intensity and 0.5 mJ per pulse incident on the sample with a radius of 25 $\mathrm{\mu}m$, consistent with experimental flux. The energy deposited at each depth z in the sample is given by the Beer Lambert law $I_0=\mu_{en}e^{-\mu z}$ where $I_0$ is is the incident X-ray intensity, $\mu_{en}$ is the linear attenuation coefficient of the material and $\mu$ is is the linear energy-absorption coefficient.  The X-ray energy deposited in the sample initially ionizes atoms, which then thermalize quickly, on the order of ps, short compared to the pulse separation of 33 ms.  For simplicity, we assume that the X-ray pulse energy goes only into heating and phase changes of the sample.  The solid Bi2Se3 starts at room temperature, and is  (1) heated to the melting temperature; (2) undergoes a phase change from solid to liquid; (3) the liquid then heats to the vaporization temperature; (4) undergoes a liquid to vapor phase change; (5) the Bi2Se3 molecule is decomposed into its atoms; (6) the free vapor atoms are heated as an ideal gas.  The spatial and temporal evolution was obtained by using the COMSOL finite element modeling package, including the enthalpy of phase change and heat capacity appropriate for each phase.

Our COMSOL model used a 0.5 ns pulse length, a compromise between the time to solve and the real pulse length.  Since thermal transport is much slower than this time scale, it does not change the longer time scale result.  The model was made using phase change to model the sample temperatures; however the volume change, mass transport, and convective effects were not included.  Surface radiation was found to have a negligible effect and was not included in the models presented here.

We use a solid density of 6800 kg/m3 for Bi2Se3, with a heat capacity of 0.190 kJ/(kg *K) and thermal conductivity of 2 W/(m*K) for the solid.  The melting temperatures taken as 980 K, the enthalpy of melting is 131 kJ/kg, and the heat capacity of 0.260 kJ/(kg*K) up to the vaporization temperature of 1600 K. The enthalpy of vaporization is 320 kJ/kg, and the enthalpy of formation of 213 kJ/kg. 

\begin{figure}
	\centering 
	\includegraphics[width=0.4\textwidth]{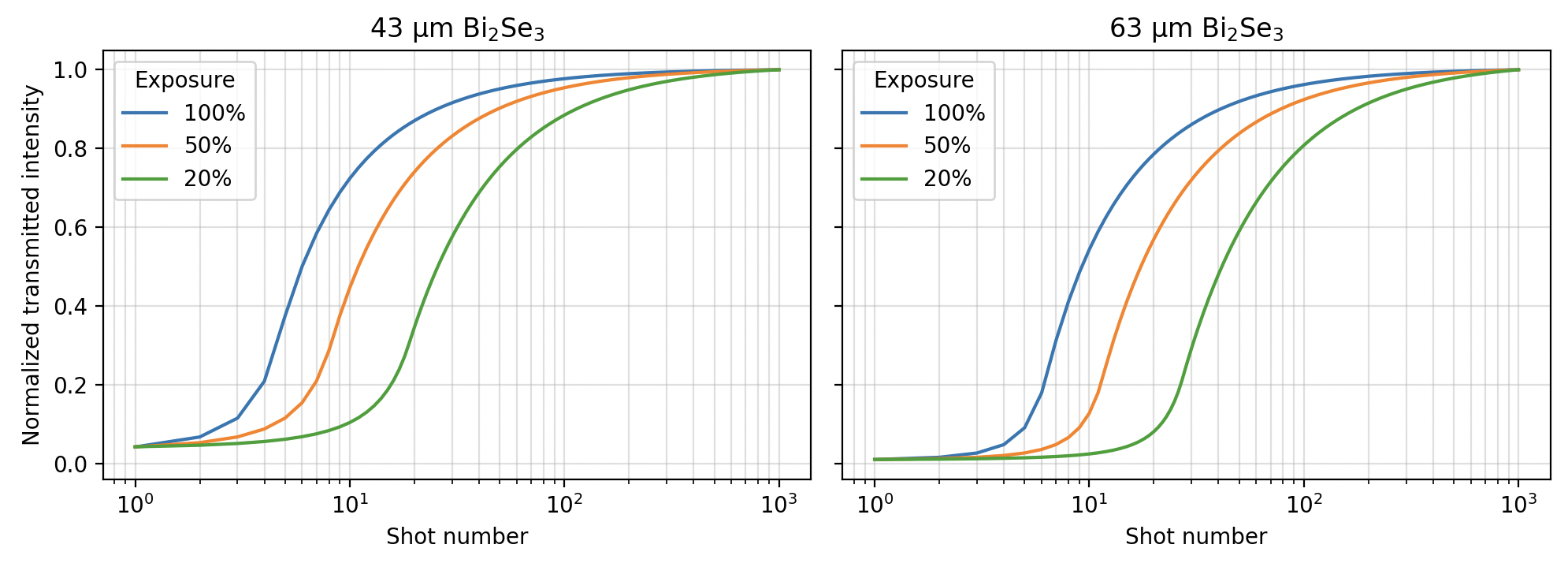}	
	\caption{Representative Beer–Lambert material-removal model illustrating the evolution of transmitted X-ray intensity during repeated XFEL irradiation for initial sample thicknesses of 43 $\mathrm{\mu}$m and 63 $\mathrm{\mu}$m as a function of exposure level.} 
	\label{fig_3A}%
\end{figure}

To illustrate the expected evolution of the transmitted X-ray signal during cumulative irradiation, we developed a simplified material-removal model based on Beer–Lambert attenuation (fig \ref{fig_3A}). A two-dimensional Gaussian beam profile is used to represent the incident XFEL intensity, and the amount of material removed after each pulse is assumed to be proportional to the local incident intensity. The remaining sample thickness is updated after every pulse, and the transmitted intensity is then calculated through the updated thickness. The model is shown for both 43 $\mathrm{\mu}$m and 63 $\mathrm{\mu}$m thick samples at multiple exposure levels and is intended as a representative simulation rather than a fit to the experimental data.
\section{Williamson-Hall}
To assess the changes to the crystalline domain size during irradiation, we performed Williamson-Hall (WH) analysis on the XRD patterns. The WH method was used to separate peak broadening contributions from crystallite size and lattice strain. This analysis provides an estimate of the crystallite size as the sample evolves from a single-crystalline diffraction pattern toward a ring-like polycrystalline diffraction under X-ray exposure.

For each XFEL pulse number, the two-dimensional X-ray diffraction image was first corrected for detector background and normalized by the measured pulse intensity. The detector geometry was calibrated using a polycrystalline CeO$_2$ standard and the known diffraction ring positions, allowing each detector pixel to be converted to scattering angle $2\theta$ and azimuthal angle $\eta$. Diffraction intensity was then integrated over the full azimuthal range of the detector ($20$--$50^\circ$), to obtain a one-dimensional intensity profile as a function of $2\theta$.

Diffraction peaks were fit using a Pearson VII peak function. The peak center, FWHM, and uncertainty were extracted for each reflection. Peaks were included in the WH fit only when they were isolated from neighboring peaks and had a signal-to-background ratio above a minimum threshold ($\mathrm{S/B} > 3$). Reflections used in the WH analysis included the Bi$_2$Se$_3$ peaks that were consistently observable and well separated in the integrated patterns.

The measured FWHM values were converted from degrees to radians and corrected for instrumental broadening according to
\begin{equation}
\beta = \sqrt{\beta_{\mathrm{meas}}^{2}-\beta_{\mathrm{inst}}^{2}}
\end{equation}
where $\beta_{\mathrm{meas}}$ is the fitted peak width and $\beta_{\mathrm{inst}}$ is the instrumental contribution to broadening. The instrumental broadening $\beta_{\mathrm{inst}}$ was determined from a reference measurement of the CeO$_2$ standard collected under identical geometry and fitting procedure and was subtracted in quadrature.

The corrected peak width $\beta$ was then used in the WH relation. 

\begin{equation}
\beta \cos\theta = \frac{K\lambda}{D} + 4\epsilon\sin\theta
\end{equation}

where $\theta$ is half the Bragg angle, $K$ is the Scherrer shape factor, $\lambda$ is the X-ray wavelength, $D$ is the coherent crystallite size, and $\epsilon$ is the microstrain. A shape factor of $K = 0.9$ was used. The X-ray wavelength was calculated from the incident photon energy of 9.7~keV ($\lambda = 1.27819$\AA).

For each pulse number and exposure condition, $\beta \cos\theta$ was plotted as a function of $4\sin\theta$. A linear fit (weighted by the peak-width uncertainties) was then used to extract the crystallite size and microstrain, where the intercept corresponds to $K\lambda/D$ and the slope corresponds to $\epsilon$. The coherent crystallite size was calculated as
\begin{equation}
D = \frac{K\lambda}{b}
\end{equation}
where $b$ is the fitted intercept.

Uncertainties in $D$ and $\epsilon$ were estimated from the standard errors of the fitted intercept and slope (propagated to $D$ by standard error propagation). The shaded error region in the WH-derived grain-size-versus-time plots represents the standard error of the mean within 1~s time bins (30 frames) for each exposure condition. Fits yielding non-physical intercepts (negative $b$) or insufficient peak counts were excluded from further analysis.
%% \label{}

%% If you have bibdatabase file and want bibtex to generate the
%% bibitems, please use
%%
\bibliographystyle{elsarticle-harv} 
\bibliography{example}

@article{Zhang2009Topological,

  title={Topological insulators in Bi2Se3, Bi2Te3 and Sb2Te3 with a single Dirac cone on the surface},

  author={Zhang, Haijun and Liu, Chao-Xing and Qi, Xiao-Liang and Dai, Xi and Fang, Zhong and Zhang, Shou-Cheng},

  journal={Nature Physics},

  volume={5},

  number={6},

  pages={438--442},

  year={2009},

  publisher={Nature Publishing Group},

  doi={10.1038/nphys1270}

}

@article{Mishra1997Thermoelectric,

  title={Electronic structure and thermoelectric properties of bismuth telluride and bismuth selenide},

  author={Mishra, S. K. and Satpathy, S. and Jepsen, O.},

  journal={Journal of Physics: Condensed Matter},

  volume={9},

  number={2},

  pages={461--470},

  year={1997},

  publisher={IOP Publishing},

  doi={10.1088/0953-8984/9/2/014}

}

@article{Das2025Bi2Se3,
  title={Enhanced Thermoelectric Performance of Bi$_2$Se$_3$-Based Materials Through Electronic Structure Engineering},
  author={Das, Soumyadeep and Chatterjee, Sayan and Roy, Souvik and Maiti, Tanusri},
  journal={Advanced Materials Technologies},
  volume={10},
  number={9},
  year={2025},
  publisher={Wiley},
  doi={10.1002/admt.202501956}
}

@article{Banerjee2025Bi2Se3,
  title={Interfacial Engineering of Bi$_2$Se$_3$-Based Heterostructures for High-Performance Thermoelectric and Energy Applications},
  author={Banerjee, Ayan and Choudhury, Anirban and Ghosh, Saptarshi and Biswas, Kanishka},
  journal={Materials Horizons},
  volume={12},
  number={10},
  pages={--},
  year={2025},
  publisher={Royal Society of Chemistry},
  doi={10.1039/D4MH01625D}
}

@article{Yin2016Bi2Se3,
  title={Bi$_2$Se$_3$ Topological Insulator Nanostructures for Near-Infrared-II Photoacoustic Imaging and Photothermal Therapy},
  author={Yin, Wei and Yu, Jie and Lv, Feng and Yan, Lei and Zheng, Lingling and Gu, Zhen and Zhao, Yanli},
  journal={ACS Nano},
  volume={10},
  number={12},
  pages={11000--11011},
  year={2016},
  publisher={American Chemical Society},
  doi={10.1021/acsnano.6b00272}
}

@article{Batool2022Barium,
  title={Synthesis, structural, optical, and dielectric properties of novel barium-doped bismuth selenide},
  author={Batool, Zainab and Atiq, Saira and Naseem, Shahzad and Mahmood, Arshad},
  journal={Materials Chemistry and Physics},
  volume={293},
  pages={126860},
  year={2023},
  publisher={Elsevier},
  doi={10.1016/j.matchemphys.2022.126860}
}

@article{Liu2024Bi2Se3,
  title={High-Performance Flexible Photodetectors Based on Bi$_2$Se$_3$ Nanosheets with Enhanced Optoelectronic Response},
  author={Liu, Yifan and Zhang, Hao and Chen, Xin and Wang, Rui and Li, Jing and Zhao, Qiang},
  journal={ACS Applied Materials \& Interfaces},
  volume={16},
  number={3},
  pages={4125--4134},
  year={2024},
  publisher={American Chemical Society},
  doi={10.1021/acsami.3c15315}
}

@article{Li2021Bi2Se3,
  title={Boosting thermoelectric performance in Bi$_2$Se$_3$-based materials through defect and interface engineering},
  author={Li, Zhihua and Wang, Yubo and Chen, Xiaolong and Zhao, Lidong},
  journal={Journal of Materials Chemistry C},
  volume={9},
  number={35},
  pages={11589--11600},
  year={2021},
  publisher={Royal Society of Chemistry},
  doi={10.1039/D1TC02613E}
}

@article{Chiatti2016,
  title={2D layered transport properties from topological insulator Bi$_2$Se$_3$ single crystals and micro flakes},
  author={Chiatti, O. and Riha, C. and Lawrenz, D. and Plech{\'a}{\v{c}}ek, T. and K{\"o}hler, A. and Hanke, M. and Steiner, P. and Lang, M.},
  journal={Scientific Reports},
  volume={6},
  pages={27483},
  year={2016},
  publisher={Nature Publishing Group},
  doi={10.1038/srep27483}
}

@article{MAZUMDER2021161492,
title = {A brief review of Bi2Se3 based topological insulator: From fundamentals to applications},
journal = {Journal of Alloys and Compounds},
volume = {888},
pages = {161492},
year = {2021},
issn = {0925-8388},
doi = {https://doi.org/10.1016/j.jallcom.2021.161492},
url = {https://www.sciencedirect.com/science/article/pii/S0925838821029017},
author = {Kushal Mazumder and Parasharam M. Shirage}
}

@article{Hong2010Ultrathin,
  title={Ultrathin Topological Insulator Bi$_2$Se$_3$ Nanoribbons Exfoliated by Atomic Force Microscopy},
  author={Hong, Seunghyun S. and Cha, Jeremy J. and Kong, Desheng and Cui, Yi},
  journal={Nano Letters},
  volume={10},
  number={8},
  pages={3118--3122},
  year={2010},
  publisher={American Chemical Society},
  doi={10.1021/nl101884h}
}

@article{Jurczyszyn2020Surface,
  title={Studies of surface states in Bi$_2$Se$_3$ induced by the BiSe substitution in the crystal subsurface structure},
  author={Jurczyszyn, Micha{\l} and Sikora, Marek and Chrobak, Maciej and Jurczyszyn, Leszek},
  journal={Applied Surface Science},
  volume={528},
  pages={146978},
  year={2020},
  publisher={Elsevier},
  doi={10.1016/j.apsusc.2020.146978}
}

@article{Hasan2020Electronic,
  title={Electronic structure of 9 quintuple layers Bi$_2$Se$_3$ within Density Functional Theory},
  author={Hasan, Md. and Hossain, Md. and Islam, M. T. and Hossain, J.},
  journal={IOP Conference Series: Materials Science and Engineering},
  volume={902},
  number={1},
  pages={012061},
  year={2020},
  publisher={IOP Publishing},
  doi={10.1088/1757-899X/902/1/012061}
}

@article{Sahu2018Weak,
  title={Weak antilocalization and low-temperature characterization of sputtered polycrystalline bismuth selenide},
  author={Sahu, Protyush and Chen, Jun Yang and Myers, Jason C. and Wang, Jian-Ping},
  journal={Applied Physics Letters},
  volume={112},
  number={12},
  pages={122402},
  year={2018},
  publisher={AIP Publishing},
  doi={10.1063/1.5020788}
}

@article{Medvedev2020ElectronPhonon,
  title={Electron-phonon coupling in metals at high electronic temperatures},
  author={Medvedev, Nikita and Milov, Igor},
  journal={Physical Review B},
  volume={102},
  number={6},
  pages={064302},
  year={2020},
  publisher={American Physical Society},
  doi={10.1103/PhysRevB.102.064302}
}

@article{Medvedev2011ShortTime,
  title={Short-Time Electron Dynamics in Aluminum Excited by Femtosecond Extreme Ultraviolet Radiation},
  author={Medvedev, Nikita and Ziaja, Beata and van der Spoel, David and others},
  journal={Physical Review Letters},
  volume={107},
  number={16},
  pages={165003},
  year={2011},
  publisher={American Physical Society},
  doi={10.1103/PhysRevLett.107.165003}
}

@article{Rethfeld2010Interaction,
  title={Interaction of dielectrics with femtosecond laser pulses: application of kinetic approach and multiple rate equation},
  author={Rethfeld, B. and Brenk, O. and Medvedev, N. and Krutsch, H.},
  journal={Applied Physics A},
  volume={101},
  number={1},
  pages={19--25},
  year={2010},
  publisher={Springer},
  doi={10.1007/s00339-010-5780-3}
}

@article{Medvedev2015Femtosecond,
  title={Femtosecond X-ray induced electron kinetics in dielectrics: application for FEL-pulse-duration monitor},
  author={Medvedev, N.},
  journal={Applied Physics B},
  volume={118},
  number={3},
  pages={417--429},
  year={2015},
  publisher={Springer},
  doi={10.1007/s00340-015-6005-4}
}

@article{PhysRevB.69.054102,
  title = {Electronic transport and consequences for material removal in ultrafast pulsed laser ablation of materials},
  author = {Bulgakova, N. M. and Stoian, R. and Rosenfeld, A. and Hertel, I. V. and Campbell, E. E. B.},
  journal = {Phys. Rev. B},
  volume = {69},
  issue = {5},
  pages = {054102},
  numpages = {12},
  year = {2004},
  month = {Feb},
  publisher = {American Physical Society},
  doi = {10.1103/PhysRevB.69.054102},
  url = {https://link.aps.org/doi/10.1103/PhysRevB.69.054102}
}

@article{Lee2011Xray,
  title={Theoretical study of x-ray absorption of three-dimensional topological insulator Bi$_2$Se$_3$},
  author={Lee, Hyun C.},
  journal={Physical Review B},
  volume={83},
  number={19},
  pages={193107},
  year={2011},
  publisher={American Physical Society},
  doi={10.1103/PhysRevB.83.193107}
}

@article{Oksengendler2024Radiation,
  title={Features of Radiation Physics of Topological Insulators},
  author={Oksengendler, B. L. and Maksimov, S. E. and Suleymanov, S. H. and Ashurov, M. H. and Nuritdinov, I. and Nikiforova, N. N. and Iskandarova, F. A. and Nuzhdov, G. S. and Karimov, Z. I. and Zatsepin, A. F.},
  journal={Journal of Surface Investigation: X-ray, Synchrotron and Neutron Techniques},
  volume={18},
  number={Suppl. 1},
  pages={S298--S301},
  year={2024},
  publisher={Pleiades Publishing},
  doi={10.1134/S1027451024702203}
}

@article{Liu2014Tuning,
  title={Tuning Dirac states by strain in the topological insulator Bi$_2$Se$_3$},
  author={Liu, Y. and Li, Y. and Rajput, S. and Gilks, D. and Lari, L. and Galindo, P. L. and Weinert, M. and Lazarov, V. K. and Li, L.},
  journal={Nature Physics},
  volume={10},
  number={4},
  pages={294--299},
  year={2014},
  publisher={Nature Publishing Group},
  doi={10.1038/nphys2898}
}

@article{10.1063/1.4975819,
    author = {Sun, Jifeng and Singh, David J.},
    title = {Using gapped topological surface states of Bi2Se3 films in a field effect transistor},
    journal = {Journal of Applied Physics},
    volume = {121},
    number = {6},
    pages = {064301},
    year = {2017},
    month = {02},
    issn = {0021-8979},
    doi = {10.1063/1.4975819},
    url = {https://doi.org/10.1063/1.4975819},
    eprint = {https://pubs.aip.org/aip/jap/article-pdf/doi/10.1063/1.4975819/15190528/064301\_1\_online.pdf},
}

@article{Dong2026Lattice,
  title={Lattice Expansion Leads to the Renormalization of Bulk-State Valence Band Structure and Band Gap in Bi$_2$Se$_3$},
  author={Dong, Jingwei and Li, Tongrui and Yu, Han and Xia, Shucai and Chen, Wei and Wen, Bo and Ren, Zefeng and Sun, Zhe and Yang, Xueming and Zhou, Chuanyao},
  journal={The Journal of Physical Chemistry Letters},
  volume={17},
  number={1},
  pages={172--177},
  year={2026},
  publisher={American Chemical Society},
  doi={10.1021/acs.jpclett.5c03534}
}

@article{Yao2022Exploring,
title={Exploring Femtosecond Laser Ablation by Snapshot Ultrafast Imaging and Molecular Dynamics Simulation},
author={Jiali Yao and D. Qi and Hongtao Liang and Yilin He and Yunhua Yao and T. Jia and Yang Yang and Zhenrong Sun and Shian Zhang},
journal={Ultrafast Science},
year={2022},
doi={10.34133/2022/9754131}
}

@article{Guo2019Ultrafast,
title={Ultrafast dynamics observation during femtosecond laser-material interaction},
author={B. Guo and Jingya Sun and Y. Lu and Lan Jiang},
journal={International Journal of Extreme Manufacturing},
year={2019},
volume={1},
doi={10.1088/2631-7990/ab3a24}
}

@article{Winter2020Ultrafast,
title={Ultrafast pump-probe ellipsometry and microscopy reveal the surface dynamics of femtosecond laser ablation of aluminium and stainless steel},
author={J. Winter and S. Rapp and Maximilian Spellauge and C. Eulenkamp and Michael Schmidt and H. Huber},
journal={Applied Surface Science},
year={2020},
volume={511},
pages={145514},
doi={10.1016/j.apsusc.2020.145514}
}

@article{Sun2024Dynamics,
title={Dynamics of nanoscale phase decomposition in laser ablation},
author={Yanwen Sun and Chaobo Chen and Thies J. Albert and Haoyuan Li and Mikhail I. Arefev and Ying Chen and Mike Dunne and J. Glownia and M. Jerman and M. Hoffmann and Matthew J. Hurley and M. Mo and Quynh L. Nguyen and Takahiro Sato and Sanghoon Song and Peihao Sun and Mark Sutton and S. Teitelbaum and Antonios S. Valavanis and Nan Wang and D. Zhu and L. Zhigilei and K. Sokolowski-Tinten},
journal={Communications Materials},
year={2024},
volume={6},
doi={10.1038/s43246-025-00785-4}
}

@article{Parsons2024Ablation,
  title={Study of ablation and shock generation across three orders of magnitude of laser intensity with 100 ps laser pulses},
  author={Parsons, S. E. and Armstrong, M. R. and Lee, H. J. and Gleason, A. E. and Goncharov, A. F. and Belof, J. and Prakapenka, V. and Granados, E. and Beg, F. N. and Radousky, H. B.},
  journal={Applied Physics Letters},
  volume={125},
  number={16},
  pages={164104},
  year={2024},
  publisher={AIP Publishing},
  doi={10.1063/5.0222979}
}

@article{Burdt2009Scaling,
  title={Experimental scaling law for mass ablation rate from a Sn plasma generated by a 1064 nm laser},
  author={Burdt, R. A. and Yuspeh, S. and Sequoia, K. L. and Tao, Y. and Tillack, M. S. and Najmabadi, F.},
  journal={Journal of Applied Physics},
  volume={106},
  number={3},
  pages={033310},
  year={2009},
  publisher={AIP Publishing},
  doi={10.1063/1.3190537}
}

@article{Joshi2023Observation,
  title={Observation of laser ablation of silicon as a function of pulse length at constant fluence via time-resolved x-ray spectroscopy},
  author={Joshi, T. R. and Bailly-Grandvaux, M. and Turner, R. E. and Spielman, R. B. and Garay, J. E. and Beg, F. N.},
  journal={Physics of Plasmas},
  volume={30},
  number={12},
  pages={122109},
  year={2023},
  publisher={AIP Publishing},
  doi={10.1063/5.0175937}
}

@book{Colvin2013Extreme,
  title={Extreme Physics: Properties and Behavior of Matter at Extreme Conditions},
  author={Colvin, Jeffrey and Larsen, John},
  edition={1st},
  year={2013},
  publisher={Cambridge University Press},
  address={Cambridge},
  isbn={9780521890665}
}

@article{Olbinado2018Ultrahigh,
  title={Ultra high-speed x-ray imaging of laser-driven shock compression using synchrotron light},
  author={Olbinado, Margie P. and Cantelli, Valentina and Mathon, Olivier and Pascarelli, Sakura and Grenzer, Joerg and Pelka, Alexander and Roedel, Melanie and Prencipe, Irene and Laso Garcia, Alejandro and Helbig, Uwe},
  journal={Journal of Physics D: Applied Physics},
  volume={51},
  number={5},
  pages={055601},
  year={2018},
  publisher={IOP Publishing},
  doi={10.1088/1361-6463/aaa2f2}
}

@article{Park2015Synchrotron,
  title={High-energy synchrotron x-ray techniques for studying irradiated materials},
  author={Park, J. S. and Zhang, X. and Sharma, H. and Kenesei, P. and Almer, J. and Li, M. and Wang, Y.},
  journal={Journal of Materials Research},
  volume={30},
  number={9},
  pages={1380--1391},
  year={2015},
  publisher={Cambridge University Press},
  doi={10.1557/jmr.2015.50}
}

@article{Yabashi2017Next,
  title={The next ten years of X-ray science},
  author={Yabashi, Makina and Tanaka, Hitoshi},
  journal={Nature Photonics},
  volume={11},
  number={1},
  pages={12--14},
  year={2017},
  publisher={Nature Publishing Group},
  doi={10.1038/nphoton.2016.251}
}

@article{Chapman2014Diffraction,
  title={Diffraction before destruction},
  author={Chapman, Henry N. and Caleman, Carl and Timneanu, Nicusor},
  journal={Philosophical Transactions of the Royal Society B: Biological Sciences},
  volume={369},
  number={1647},
  pages={20130313},
  year={2014},
  publisher={The Royal Society},
  doi={10.1098/rstb.2013.0313}
}

@article{Hull2019Early,
  title={Early time dynamics of laser-ablated silicon using ultrafast grazing incidence X-ray scattering},
  author={Hull, C. and Raj, S. and Lam, R. and Katayama, T. and Pascal, T. and Drisdell, W. S. and Saykally, R. and Schwartz, C. P.},
  journal={Chemical Physics Letters},
  volume={736},
  pages={136811},
  year={2019},
  publisher={Elsevier},
  issn={0009-2614},
  doi={10.1016/j.cplett.2019.136811}
}

@article{Randolph2022Nanoscale,
  title={Nanoscale subsurface dynamics of solids upon high-intensity femtosecond laser irradiation observed by grazing-incidence x-ray scattering},
  author={Randolph, Lisa and Banjafar, Mohammadreza and Preston, Thomas R. and Yabuuchi, Toshinori and Makita, Mikako and Dover, Nicholas P. and R{\"o}del, Christian and G{\"o}de, Sebastian and Inubushi, Yuichi and others},
  journal={Physical Review Research},
  volume={4},
  number={3},
  pages={033038},
  year={2022},
  publisher={American Physical Society},
  doi={10.1103/PhysRevResearch.4.033038}
}

@article{Katagiri2025Xray,
  title={X-ray induced grain boundary formation and grain rotation in Bi$_2$Se$_3$},
  author={Katagiri, Kento and Kozioziemski, Bernard and Folsom, Eric and G{\"o}de, Sebastian and Wang, Yifan and Appel, Karen and Chalise, Darshan and Cook, Philip K. and Eggert, Jon and Howard, Marylesa and Kim, Sungwon and Kon{\^o}pkov{\'a}, Zuzana and Makita, Mikako and Nakatsutsumi, Motoaki and Nielsen, Martin M. and Pelka, Alexander and Poulsen, Henning F. and Preston, Thomas R. and Reddy, Tharun and Schwinkendorf, Jan-Patrick and Seiboth, Frank and Simons, Hugh and Wang, Bihan and Yang, Wenge and Zastrau, Ulf and Kim, Hyunjung and Dresselhaus-Marais, Leora E.},
  journal={Scripta Materialia},
  volume={256},
  pages={116416},
  year={2025},
  publisher={Elsevier},
  issn={1359-6462},
  doi={10.1016/j.scriptamat.2024.116416}
}

@article{Kang2017Hard,
  title={Hard X-ray free-electron laser with femtosecond-scale timing jitter},
  author={Kang, Heung-Sik and Min, Chang-Ki and Heo, Hoon and Kim, Changbum and Yang, Haeryong and Kim, Gyujin and Nam, Inhyuk and Baek, Soung Youl and Choi, Hyo-Jin and Mun, Geonyeong and Park, Byoung Ryul and Suh, Young Jin and Shin, Dong Cheol and Hu, Jinyul and Hong, Juho and Jung, Seonghoon and Kim, Sang-Hee and Kim, KwangHoon and Na, Donghyun and Park, Soung Soo and Park, Yong Jung and Han, Jang-Hui and Jung, Young Gyu and Jeong, Seong Hun and Ko, In Soo},
  journal={Nature Photonics},
  volume={11},
  number={11},
  pages={708--713},
  year={2017},
  publisher={Nature Publishing Group},
  doi={10.1038/s41566-017-0013-8}
}

@article{AlonsoMori2020Femtosecond,
  title={Femtosecond electronic structure response to high intensity XFEL pulses probed by iron X-ray emission spectroscopy},
  author={Alonso-Mori, Roberto and Sokaras, Dimosthenis and Cammarata, Marco and Zhu, Diling and Chollet, Matthieu and Nelson, Stephen and Schafer, Donald W. and Kroll, Thomas and Weng, Tsu-Chien and Nordlund, Dennis and others},
  journal={Scientific Reports},
  volume={10},
  number={1},
  pages={16837},
  year={2020},
  publisher={Nature Publishing Group},
  doi={10.1038/s41598-020-74003-1}
}

@article{Tachibana2015Nanoplasma,
  title={Nanoplasma Formation by High Intensity Hard X-rays},
  author={Tachibana, T. and Jurek, Z. and Fukuzawa, H. and Motomura, K. and Kumagai, Y. and Tachibana, K. and Tono, K. and Inubushi, Y. and Yabashi, M. and Ishikawa, T. and others},
  journal={Scientific Reports},
  volume={5},
  pages={10977},
  year={2015},
  publisher={Nature Publishing Group},
  doi={10.1038/srep10977}
}

@article{Sawada2024Spatiotemporal,
  title={Spatiotemporal dynamics of fast electron heating in solid-density matter via XFEL},
  author={Sawada, H. and Yabuuchi, T. and Higashi, N. and Inubushi, Y. and Yabashi, M. and Tanaka, H. and others},
  journal={Nature Communications},
  volume={15},
  number={1},
  pages={7528},
  year={2024},
  publisher={Nature Publishing Group},
  doi={10.1038/s41467-024-51084-4}
}

@article{Sacksteder2015Modification,
  title={Modification and Control of Topological Insulator Surface States Using Surface Disorder},
  author={Sacksteder, Vincent and Ohtsuki, Tomi and Kobayashi, Koji},
  journal={Physical Review Applied},
  volume={3},
  number={6},
  pages={064006},
  year={2015},
  publisher={American Physical Society},
  doi={10.1103/PhysRevApplied.3.064006}
}

@article{Kim2018Focusing,
  title={Focusing X-ray free-electron laser pulses using Kirkpatrick-Baez mirrors at the NCI hutch of the PAL-XFEL},
  author={Kim, J. and Kim, H. Y. and Park, J. and Kim, S. and Kim, S. and Rah, S. and Lim, J. and Nam, K. H.},
  journal={Journal of Synchrotron Radiation},
  volume={25},
  number={1},
  pages={289--292},
  year={2018},
  publisher={International Union of Crystallography},
  doi={10.1107/S1600577517016186},
  pmid={29271778},
  pmcid={PMC5741134}
}

%% else use the following coding to input the bibitems directly in the
%% TeX file.

%%\begin{thebibliography}{00}

%% \bibitem[Author(year)]{label}
%% For example:

%% \bibitem[Aladro et al.(2015)]{Aladro15} Aladro, R., Martín, S., Riquelme, D., et al. 2015, \aas, 579, A101

%%\end{thebibliography}

\end{document}